\pdfoutput=1

\documentclass[final,times,1p]{elsarticle}

\usepackage{graphicx}
\usepackage{float}
\usepackage{url}

\usepackage{setspace}

\usepackage{amssymb}
\usepackage{amsmath}

\journal{Proceedings of the Combustion Institute}

\begin{document}
\begin{frontmatter}


\title{Chapman-Jouguet deflagrations and their transition to detonation}
\author[add1]{Mohamed Saif}
\author[add1]{Wentian Wang}
\author[add2]{Andrzej Pekalski}
\author[add3]{Marc Levin}
\author[add1]{Matei I. Radulescu}
\ead{matei@uottawa.ca}

\address[add1]{Department of Mechanical Engineering, University of Ottawa, Ottawa Canada K1N6N5}
\address[add2]{Shell Global Solutions (UK)}
\address[add3]{Shell Exploration \& Production Company (SEPCO)}

\begin{abstract}
We study experimentally fast flames and their transition to detonation in mixtures of methane, ethane, ethylene, acetylene, and propane mixtures with oxygen.  Following the interaction of a detonation wave with a column of cylinders of varying blockage ratio, the experiments demonstrate that the fast flames established are Chapman-Jouguet deflagrations, in excellent agreement with the self-similar model of Radulescu et al.\ \cite{Radulescuetal2015}.  The experiments indicate that these Chapman-Jouguet deflagrations dynamically restructure and amplify into fewer stronger modes until the eventual transition to detonation.  The transition length to a self-sustained detonation was found to correlate very well with the mixtures' sensitivity to temperature fluctuations, reflected by the $\chi$ parameter introduced by Radulescu, which is the product of the non-dimensional activation energy $E_a/RT$ and the ratio of chemical induction to reaction time $t_i/t_r$.  Correlation of the measured DDT lengths determined that the relevant characteristic time scale from chemical kinetics controlling DDT is the energy release or excitation time $t_r$. Correlations with the cell size also capture the dependence of the DDT length on $\chi$ for fixed blockage ratios. 
\end{abstract}


\end{frontmatter}
\section{Introduction}
\addvspace{10pt}
The accidental release of a flammable mixture (for example, in a processing plant or during transport) may lead to the formation of a flammable gas cloud. If the flammable cloud ignites, an explosion can occur, especially if the cloud envelops a congested area. When a reactive cloud is ignited, an initially low speed subsonic flame accelerates owing to various instability mechanisms and possible interactions with confining structures and obstacles \cite{Lee&Moen1980, Ciccarelli&Dorofeev2008}. The early stages of this subsonic flame acceleration are relatively well understood and modeled. The acceleration of this subsonic flame culminates with the establishment of a wave propagating close to the sound speed in the burned products with respect to them, called a choked flame \cite{Lee2008}. The subsequent transition of this choked flame into a detonation is much less understood. The present study addresses the acceleration of such high speed flames to detonations experimentally.  

The experimental configuration used to isolate these high speed waves, without relying on a prior low speed flame acceleration, is generally obtained through the interaction of a detonation wave with a porous plate \cite{Chao2006, Grondin&Lee2010, Maleyetal2015}, from which a transmitted shock - turbulent flame complex can emerge. Extending previous models of Chao \cite{Chao2006}, Radulescu et al.\ formulated a closed form self-similar model to predict these high speed deflagrations \cite{Radulescuetal2015}.  The model assumed that the rear of the reaction zone develops a limiting characteristic, such that the flow behind the reaction zone was sonic in relation to the non-reacted gas driven forward ahead of the reaction zone, i.e., the flame is a CJ deflagration. The model's predictions were found in good agreement with the experiments in methane-oxygen mixtures, suggesting that the turbulent fast flames organize gasdynamically into Chapman-Jouguet detonations prior to their transition to detonation.   In the present study, we first wish to verify whether this is a necessary condition for DDT in other reactive mixtures. Hence, the experiments are thus extended to ethane, ethylene, acetylene, and propane with oxygen as oxidizer, sometimes diluted by an argon.   

The primary focus of the present study, however, is to monitor the amplification of the fast deflagration waves into a detonation in all the mixtures studied and seek whether a universal correlation can be established.  In the past, it was argued qualitatively that the transition length from choked flames to detonations was related to the problem of gradient initiation of detonation, where a unique length scale provided the characteristic length and time scales for DDT \cite{Khokhlovetal1997, Dorofeevetal2000}.  Dorofeev et al.\ further suggested that this unique length scale can be used as the characteristic scale of obstacle spacing in a congested geometry and showed that the critical scale for detonation initiation can be correlated with the detonation cell size \cite{Dorofeevetal2000}, yielding to the called 7 $\lambda$ criterion. Nevertheless, cell sizes are very difficult to measure and only available in limited reacting mixtures and operating conditions \cite{Kaneshige&Shepherd1997}.  There is currently a need to predict or relate this critical length scale to the thermo-chemical properties of the mixture. 

In the present study, we seek whether the amplification length of Chapman-Jouguet deflagrations to detonations can be correlated to the thermo-chemical properties of the reactive mixtures.  Recent experiments performed in methane-oxygen mixtures by Maley et al. \cite{Maleyetal2015} revealed that the propagation mechanism of the choked flames is via punctuated hot spot ignitions followed by turbulent mixing.  Kuznetsov et al.\ \cite{Kuznetsovetal2000}, Chao \cite{Chao2006} and Grondin and Lee \cite{Grondin&Lee2010} also noted differences between mixtures characterized by weakly or strongly dependent reaction rates to temperature fluctuations, further suggesting that the propensity of a mixture to develop hot spots from local perturbations is essential for a rapid DDT.  Recently, Radulescu et al.\ proposed a non-dimensional parameter to characterize the detonability of reactive mixtures, namely,
\begin{align}
\chi=\frac{E_a}{RT} \frac{t_i}{t_r}\label{eq:chi}
\end{align}
where $E_a/RT$ is the reduced activation energy, $t_i$ is the ignition delay time and $t_r$ is the reaction time \cite{Radulescuetal2013}. This parameter has also been suggested to control the propensity for hot spot formation in the presence of temperature fluctuations, detonation stability, the propensity for engine knock, and the local acceleration of one-dimensional fast flames to detonations \cite{Short&Sharpe2003, Radulescu2003, Tang&Radulescu2013, Radulescuetal2013}.  The present study examines whether a correlation can be drawn between the DDT length in the various mixtures studied and this parameter. 
\section{Experimental details}
\begin{figure*}
\begin{center}
\includegraphics[width=120mm]{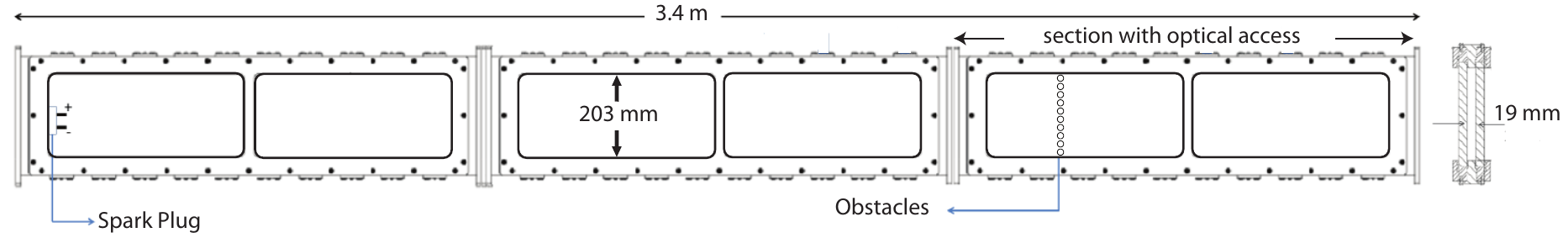}
\end{center}
\caption{Schematic of experimental set-up.}
\label{fig:schematic}
\end{figure*}

\addvspace{10pt}
The experiments were conducted in a 3.5-m-long thin rectangular channel, 203-mm-tall and 19-mm-wide, as described by Bhattacharjee \cite{Bhattacharjee2013}. A schematic is shown in Fig.\ \ref{fig:schematic}. A large-scale Edgerton shadowgraph technique \cite{Hargather&Settles2009} was implemented using a 2-m by 2-m retro-reflective screen using a high speed camera and a Xenon arc continuous light source \cite{Dennisetal2014}. A row of cylindrical obstacles of 16-mm-diameter were placed at the entrance of the visual section of the shock tube to allow for visualization of the fast flames established downstream.  Two blockage ratios were investigated, 75\% and 90\%, which permitted to change the speed of Chapman-Jouguet deflagration waves (in the laboratory frame) and the strength of the leading shock \cite{Radulescuetal2015}.  
We studied stoichiometric methane, ethane, ethylene, acetylene, and propane mixtures with oxygen as oxidizer.  The acetylene mixture was diluted with 75\% argon.  The gases were mixed in a separate vessel and left to mix for a minimum of 24 hours before an experiment. Varying the initial pressure, $p_0$, of the test mixture permitted us to control the reactivity of the mixture. The channel was evacuated to below 80 Pa absolute pressure before filling with the test mixture. A high voltage spark obtained from a capacitor discharge (1 kJ with deposition time of $2 \mu$s) was used to ignite the test mixtures.  
\section{Results}
\begin{figure*}
\begin{center}
\includegraphics[width=120mm]{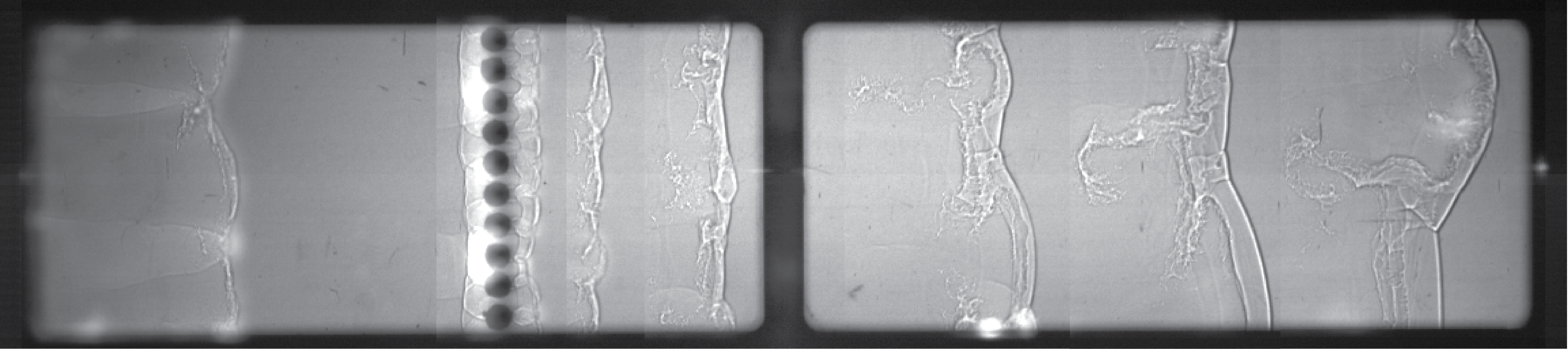}
\end{center}
\caption{Overlay of multiple shadowgraph video frames showing the evolution of the fast flame in a CH$_4$ + 2O$_2$ mixture at 8.2 kPa with a 75 \% blockage ratio; see video \ref{fig:methane75_8.2} as supplemental material.}
\label{fig:methane75_8.2}
\end{figure*}

\begin{figure*}
\begin{center}
\includegraphics[width=120mm]{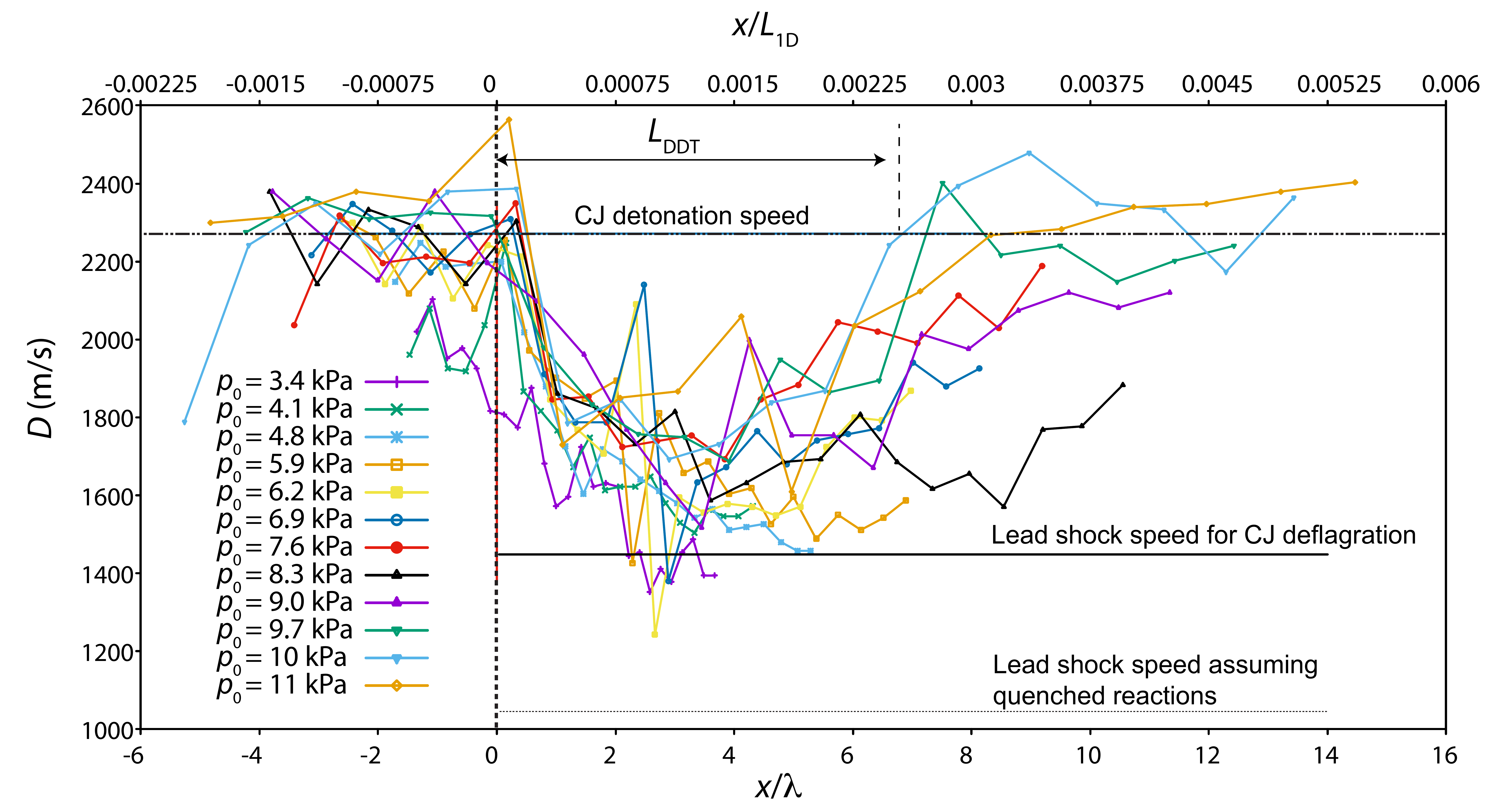}
\end{center}
\caption{Lead shock speed variation with the distance traveled from the location of the column of cylinders for the experiments conducted in CH$_4$ + 2O$_2$ using the 75 \% blockage ratio.}
\label{fig:speedsmethane75}
\end{figure*}

\begin{figure}
\begin{center}
\includegraphics[width=100mm]{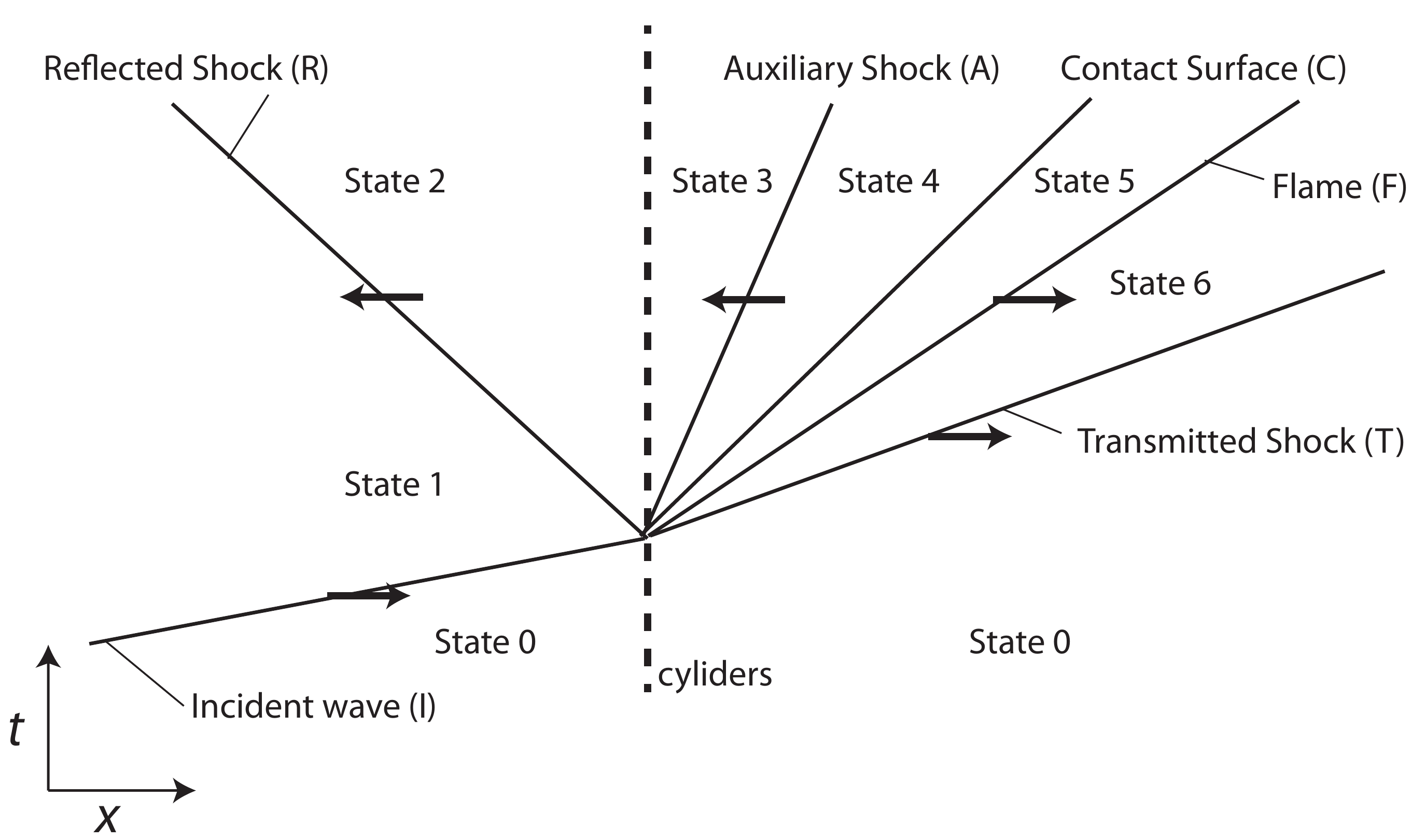}
\end{center}
\caption{Self-similar interaction model of an incident detonation partially reflecting on a finite blockage plane and the transmitted CJ deflagration supporting a forward and rear facing shock, from \cite{Radulescuetal2015}.}
\label{fig:similarmodel}
\end{figure}

\addvspace{10pt}
Figure \ref{fig:methane75_8.2} shows a multi-frame overlay of a typical test, here in stoichiometric methane-oxygen at $p_0= $8.2 kPa initial pressure for the 75\% blockage ratio configuration. The wave front is propagating from left to right.  The corresponding video \ref{fig:methane75_8.2} (supplemental material) shows the evolution of the flow field in more detail.  A cellular detonation impinges onto the row of cylinders.  Following the interaction, an irregular shock/reaction zone structure is observed, with local explosion centers associated with Mach reflections.  The details of such interactions at small scales were investigated by Maley et al. \cite{Maleyetal2015}.  As the front evolves, it organizes itself into fewer modes.  At the end of the channel, only two main triple points are observed, with local DDT occurring near the top wall from a sufficiently strong shock reflection. The structure of the fast flame resembles quite closely the structure of unstable cellular detonations \cite{Radulescuetal2007}, but propagates at an average speed approximately 35\% lower than the CJ speed, as shown in Fig. \ref{fig:speedsmethane75}. The average speed reported is the local average speed.  At a given time, the position of the lead shock was registered at five equally spaced locations along the shock, then averaged in order to obtain the mean position of the shock. The trajectory of this average shock location was used to obtain its local speed.\\
Figure \ref{fig:speedsmethane75} shows the lead shock speeds recorded for all the tests performed in methane-oxygen mixtures in the 75\% blockage ratio configuration.  After the interaction with the obstacles, the shock-flame complex propagates at approximately 65\% of the CJ detonation speed, in good agreement with the CJ deflagration speed calculated by Radulescu et al. \cite{Radulescuetal2015}.\\
The details of the model and its numerical verification can be found elsewhere \cite{Radulescuetal2015}.  Figure \ref{fig:similarmodel} illustrates its construction.  The incident detonation onto the column of cylinders yields a reflected (R) and transmitted (T) shock.  Behind the transmitted shock is a deflagration (F), assumed to be of CJ type with sonic outflow.  The matching of pressure and speeds at the contact surface (C) separating the gases expanding isentropically across the cylinder bank and the gases processed by the deflagration require an auxiliary shock (A).  Regions 0, 1, 4, 5 and 6 are uniform while the isentropic expansion from state 2 to 3 is steady.  Applying the Rankine-Hugoniot equations across each discontinuity and the steady isentropic relations between regions 2 and 3 permits to determine a unique solution.  The calculations were performed using the perfect gas relations with different specific heat ratios for the reacted and unreacted gases\cite{Radulescuetal2015}.\\
In the experiments, we find that the fast flames subsequently accelerate to a detonation.  The distance traveled by the lead shock from the column of cylinders to where it reaches the CJ detonation speed DDT is labeled $L_{DDT}$, as shown in Fig. \ref{fig:speedsmethane75}.  When normalized by the detonation cell size $\lambda$, obtained from Shepherd's detonation database \cite{Kaneshige&Shepherd1997}, an approximate correlation of $L_{DDT}\simeq 7\lambda$ can be deduced for the range of pressures investigated, in good agreement with the length recommended by Dorofeev et al. \cite{Dorofeevetal2000}.

\begin{figure}
\begin{center}
\includegraphics[width=80mm]{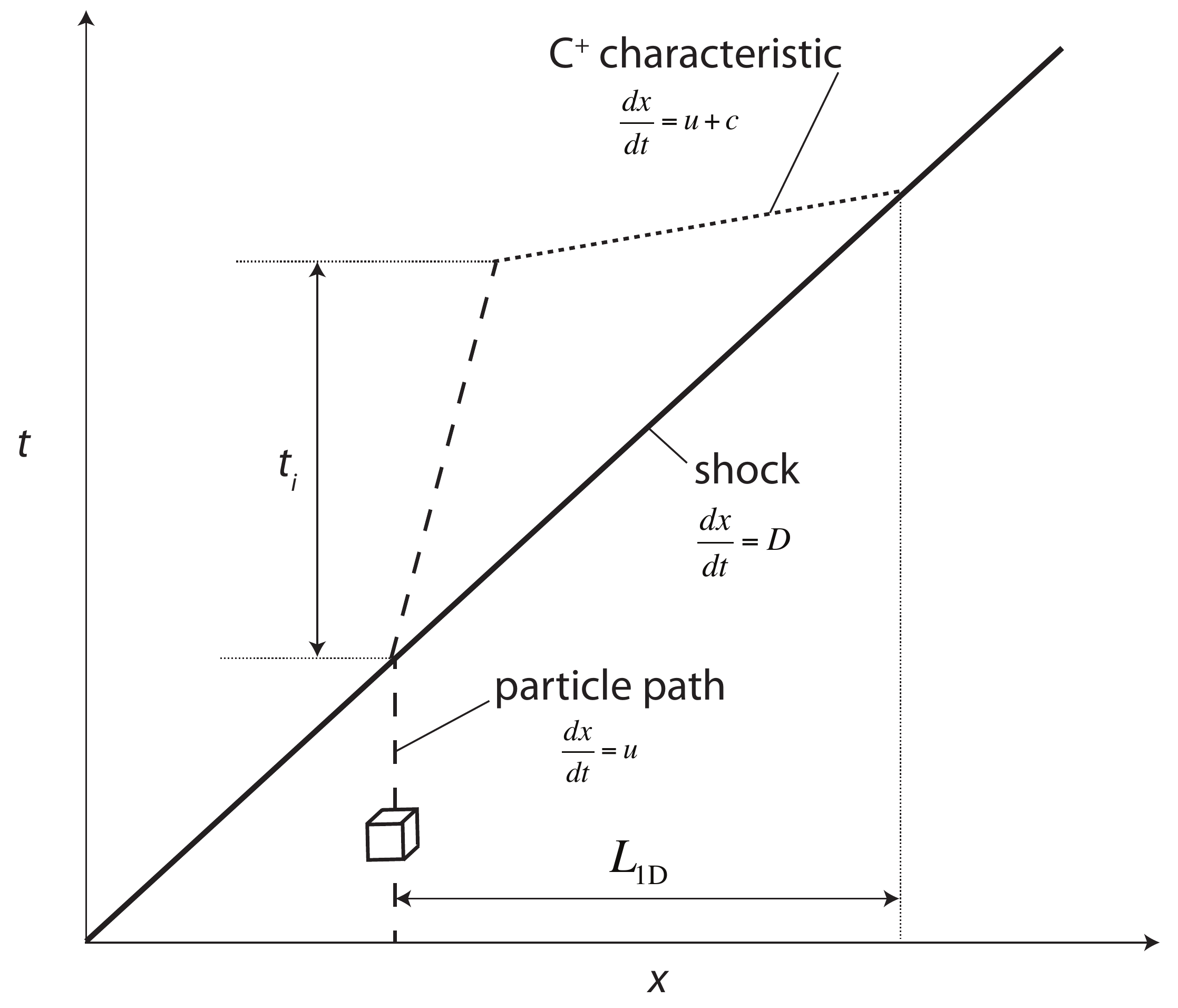}
\end{center}
\caption{Definition of the characteristic DDT length $L_{1D}$ as the distance traveled by a shock moving at speed D such that a shocked particle ignition after a delay $t_i$ communicates its gasdynamic effect to the shock along a forward facing C$^+$ characteristic.}
\label{fig:L1Ddefinition}
\end{figure}

\begin{figure*}
\begin{center}
\includegraphics[width=120mm]{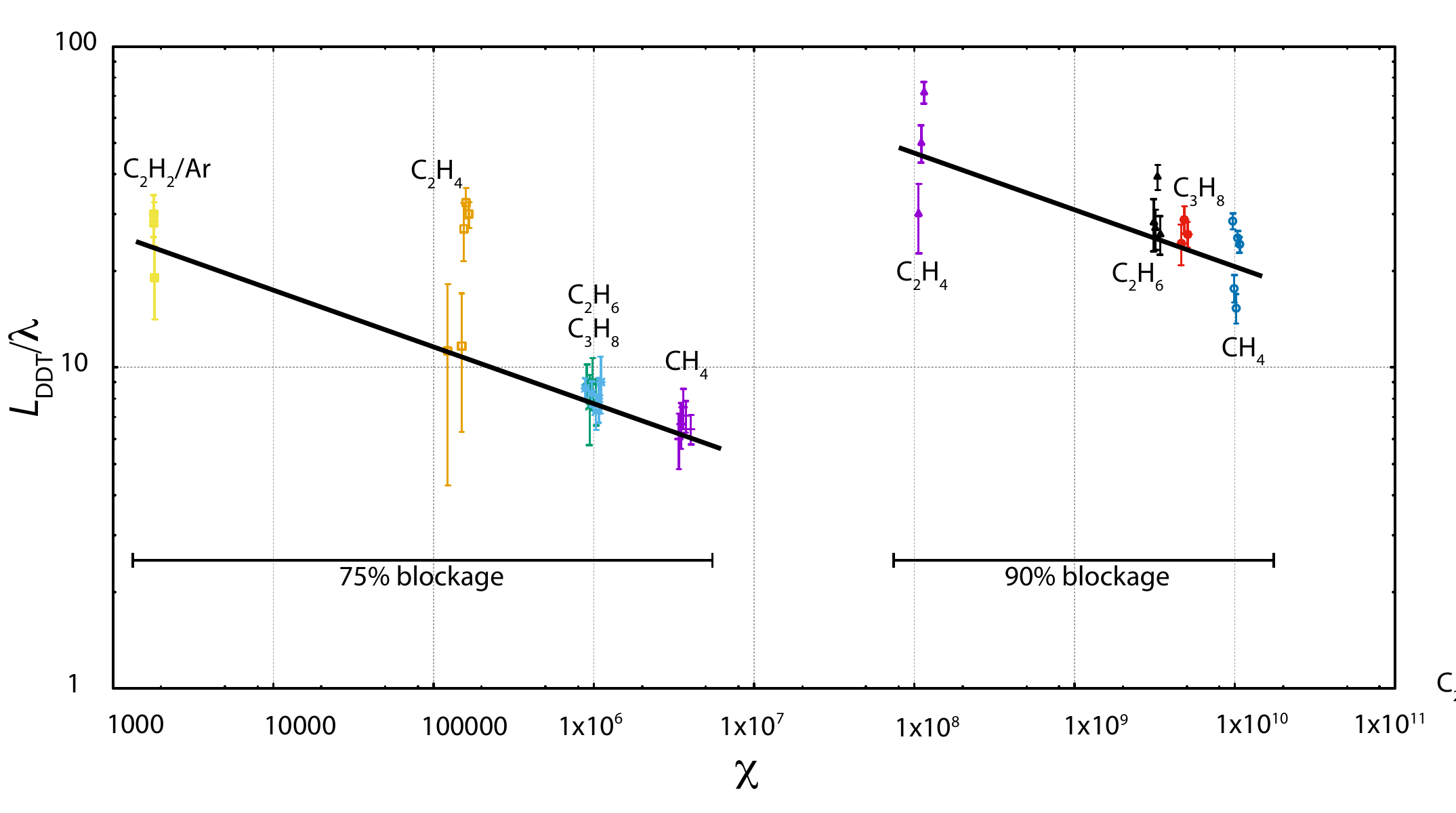}
\end{center}
\caption{Variation of the DDT length $L_{DDT}$ normalized by the mixture cell size for all experiments performed, in terms of the local hotspot parameter $\chi=(E_a/RT)(t_i/t_r)$.}
\label{fig:Lddtcellsize}
\end{figure*}

\begin{figure*}
\begin{center}
\includegraphics[width=120mm]{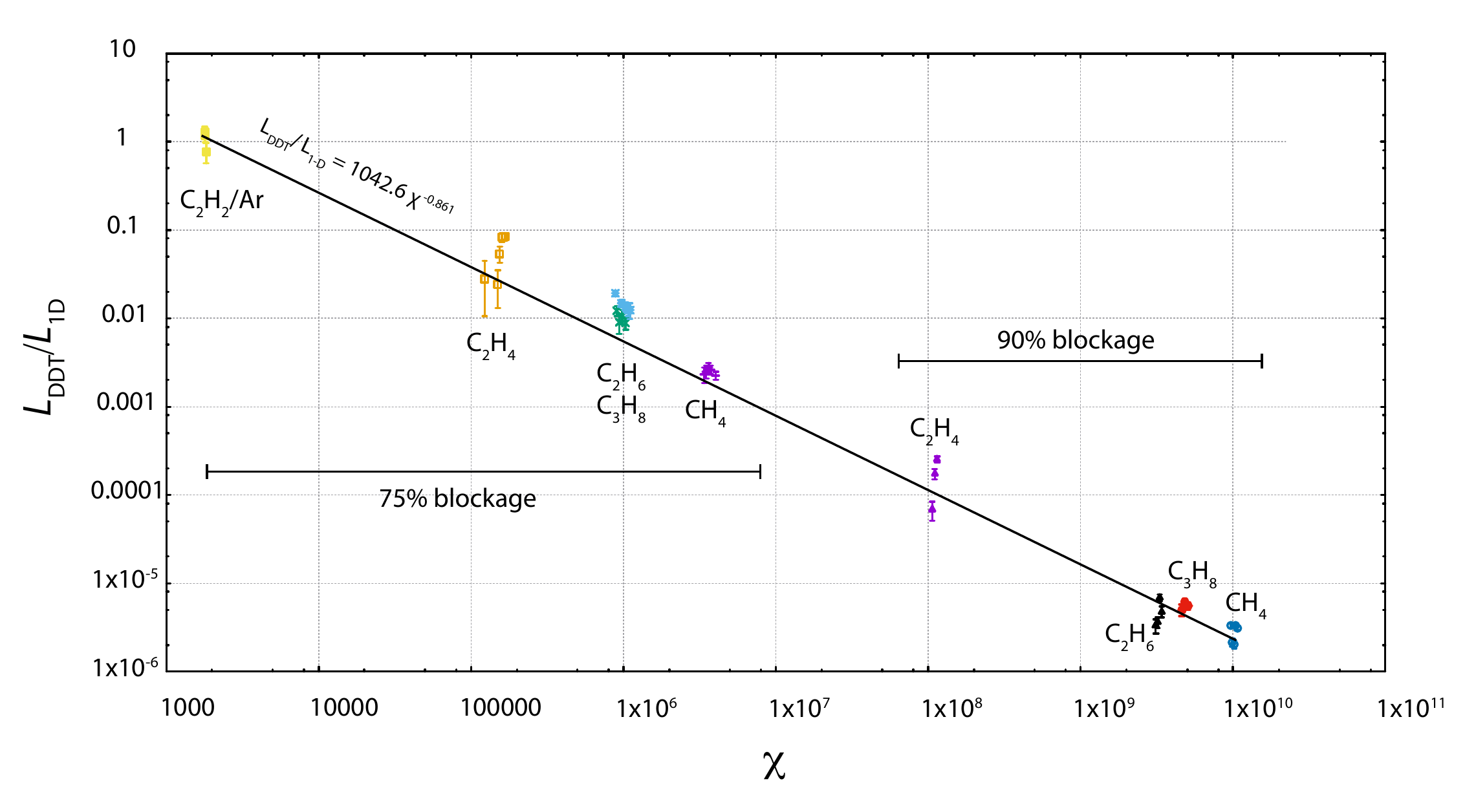}
\end{center}
\caption{Variation of the DDT length $L_{DDT}$ normalized by the one-dimensional characteristic length $L_{1D}$ defined in \eqref{eq:L1D} for all experiments performed, in terms of the local hotspot parameter $\chi=(E_a/RT)(t_i/t_r)$.}
\label{fig:LddtL1D}
\end{figure*}

\begin{figure*}
\begin{center}
\includegraphics[width=120mm]{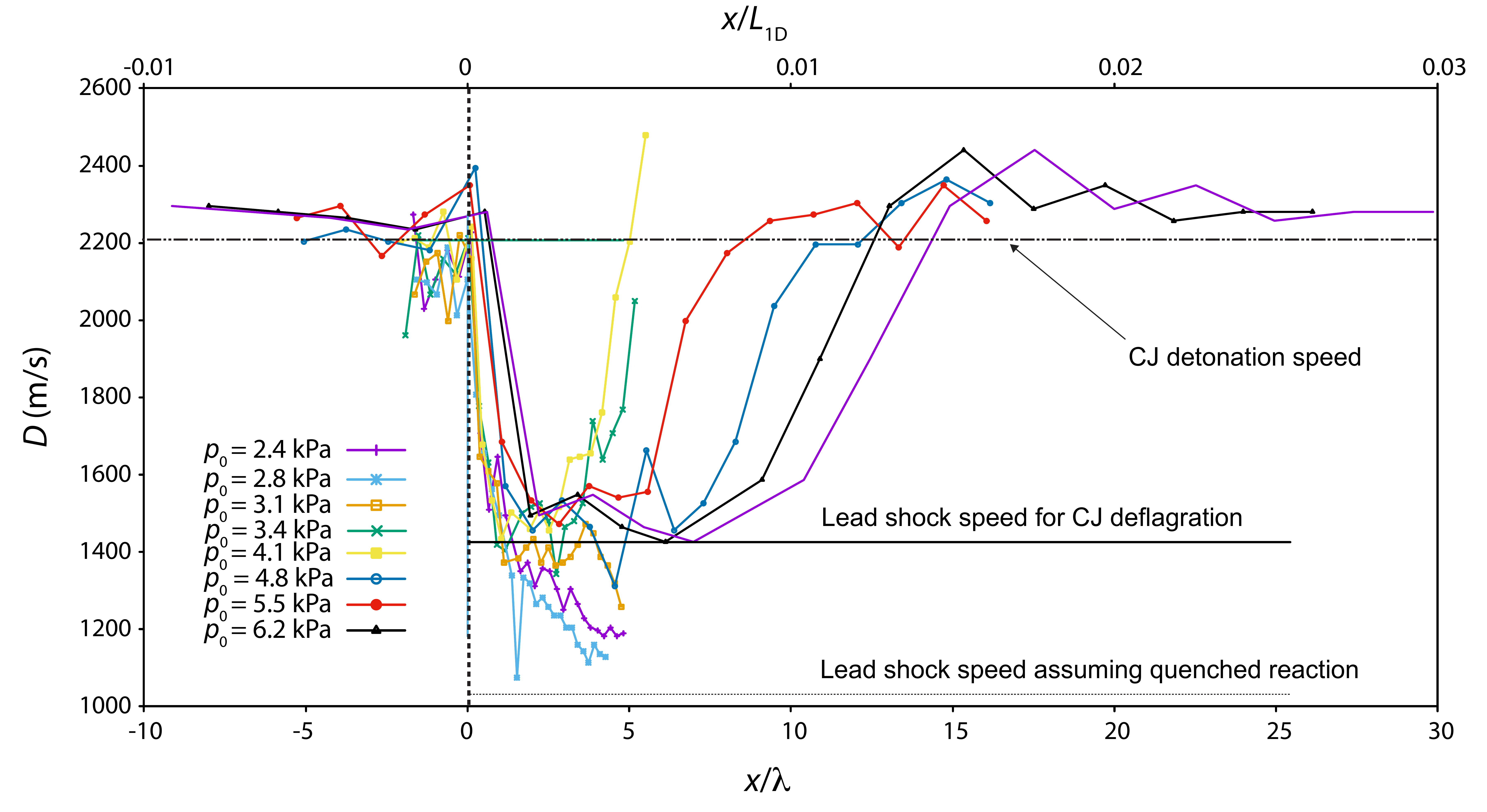}
\end{center}
\caption{Lead shock speed variation with the distance traveled from the location of the column of cylinders for the experiments conducted in C$_2$H$_6$+7O$_2$ using the 75 \% blockage ratio.}
\label{fig:speedsethane75}
\end{figure*}

We also compared the DDT length with a characteristic length $L_{1D}$ that a one-dimensional steady shock traveling at a speed $D$ compatible with the assumed Chapman Jouguet deflagration would travel before transiting to a detonation. The definition of this length scale is shown in Fig. \ref{fig:L1Ddefinition}.  This distance is the distance traveled by the lead shock during the time a particle crossing the shock ignites after a delay $t_i$ and communicates its effect to the lead shock along a forward facing $u+c$ characteristic, yielding 
\begin{align}
L_{1D}=ct_i\left(\frac{u+c}{D}-1\right)^{-1}\label{eq:L1D}
\end{align}
The state behind the shock was computed using the CEA code \cite{Gordon&McBride}, while the ignition delay $t_i$ was evaluated at constant density and internal energy using CANTERA \cite{Cantera} and the GRI-3.0 mechanism \cite{GRI3}. The DDT length was found to also be relatively invariant for all the tests performed in this mixture and blockage ratio, yielding approximately $L_{DDT}\simeq 0.003 L_{1D}$. That this ratio is significantly less than unity clearly highlights the important role played by local hotspots, permitting gas ignition much quicker than in the absence of perturbations, hence shortening the DDT length.\\
The experiments performed with the 90\% blockage ratio with the methane mixtures yielded similar results. The speed of the shock prior to DDT was in good agreement with the CJ deflagration model, with larger velocity deficits of approximately 50\% in this case of higher blockage.  The DDT length was correspondingly longer, yielding  $L_{DDT}\simeq 3x10^{-6} L_{1D}\simeq 20 \lambda$.  Clearly, the blockage ratio controls the DDT length, as discussed below in view of the experiments in all the mixtures investigated, shown in Figs. \ref{fig:Lddtcellsize} and \ref{fig:LddtL1D}.\\
Figure \ref{fig:speedsethane75} shows the results of the tests conducted for ethane at initial pressures $p_0$ ranging between 2.4 and 6.2 kPa. After the interaction with the cylinders, the lead shock speed dropped close to the computed CJ deflagration speed and continued for a distance about 5 to 10 $\lambda$ before transition.  The DDT lengths for all the ethane-oxygen mixtures are reported in Figs. \ref{fig:Lddtcellsize} and \ref{fig:LddtL1D}. The calculations were performed with the Sandiego mechanism \cite{Sandiego}.

\begin{figure*}
\begin{center}
\includegraphics[width=120mm]{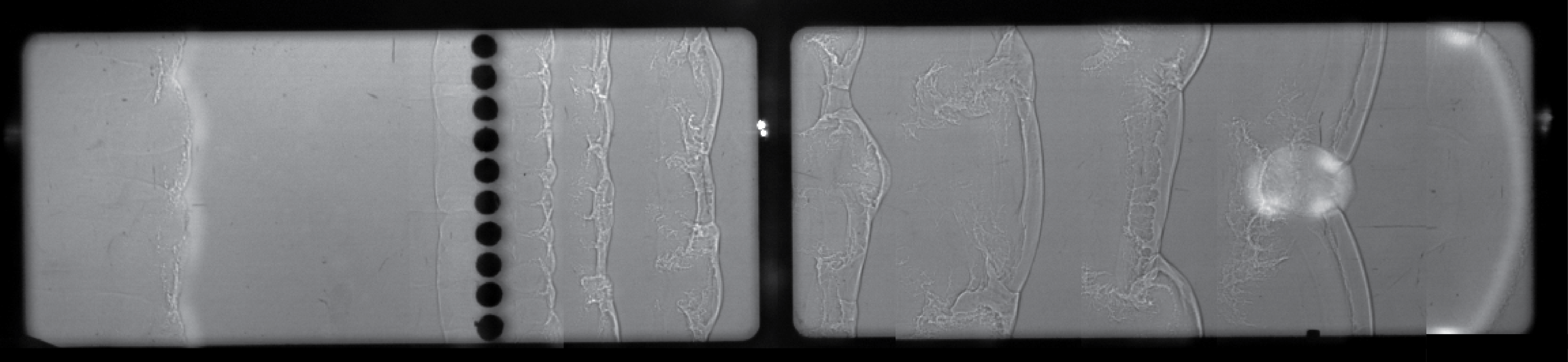}
\end{center}
\caption{Overlay of multiple shadowgraph video frames showing the evolution of the fast flame in a 2C$_2$H$_6$+ 7O$_2$ mixture at 3.4 kPa with a 75 \% blockage ratio; see video \ref{fig:ethane75_3.4}.}
\label{fig:ethane75_3.4}
\end{figure*}

An example of the DDT process in this mixture is shown in Fig. \ref{fig:ethane75_3.4} (and video \ref{fig:ethane75_3.4} as supplemental material) for the 75\% blockage ratio.  Again, the DDT takes the same form as in the methane mixture above, with the organization of the front into fewer stronger modes before the final detonation initiation in the last two frames.   

For tests conducted below 3 kPa, the speed of the shock decayed well below the predicted CJ deflagration speed.  The high speed videos indicated the decoupling between the shock and reaction zone.  This behavior suggested the growing role of losses to the narrow channel walls at lower pressures.  These effects were not further explored in the present study.

%
The tests performed in the propane and ethylene mixtures showed the same characteristics as above.  The fast flames, prior to the amplification to a detonation, propagated as CJ deflagrations.  In all cases, the front organized into fewer stronger modes prior to the final transition to a detonation.  The quantitative results for the run-up-distance as compared to the detonation cell size or the 1D run up distance are presented in Figs. \ref{fig:Lddtcellsize} and \ref{fig:LddtL1D}.  The calculations were performed using the Sandiego mechanism \cite{Sandiego}.  
 
\begin{figure*}
\begin{center}
\includegraphics[width=120mm]{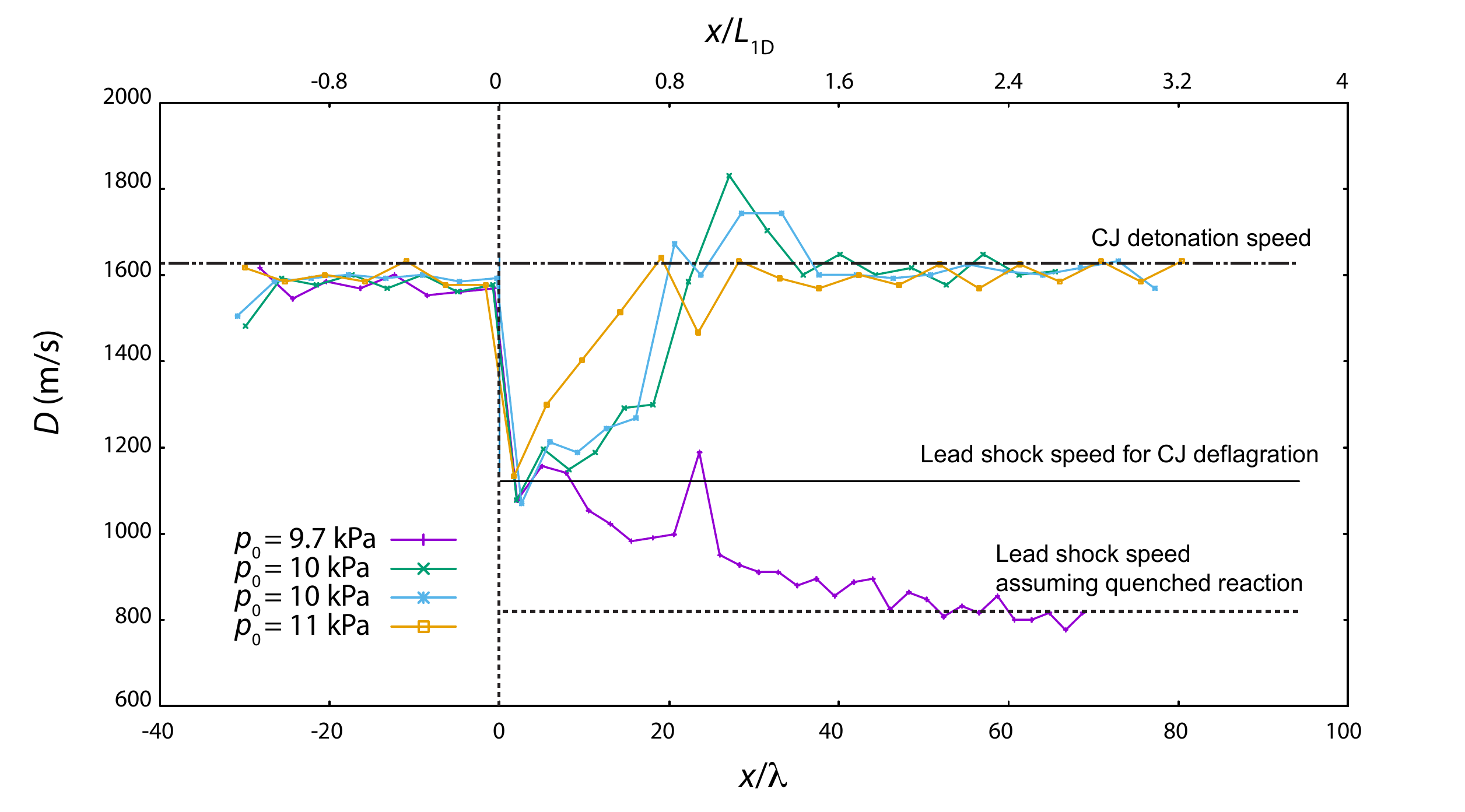}
\end{center}
\caption{Lead shock speed variation with the distance traveled from the location of the column of cylinders for the experiments conducted in 2C$_2$H$_2$ + 5O$_2$ + 21 Ar using the 75 \% blockage ratio.}
\label{fig:speedsacetylene75}
\end{figure*}

Figure \ref{fig:speedsacetylene75} shows the results for the tests conducted in acetylene-oxygen diluted with 75\% argon gas, a test mixture previously investigated by Grondin and Lee \cite{Grondin&Lee2010}.  For the re-initiation cases, a similar behavior was observed as above, with the fast flame establishing near the CJ deflagration speed, prior to its transition to a detonation.  The quantitative results for the run-up-distance as compared to the detonation cell size or the 1D run up distance are presented in Figs. \ref{fig:Lddtcellsize} and \ref{fig:LddtL1D}. 

\begin{figure*}
\begin{center}
\includegraphics[width=120mm]{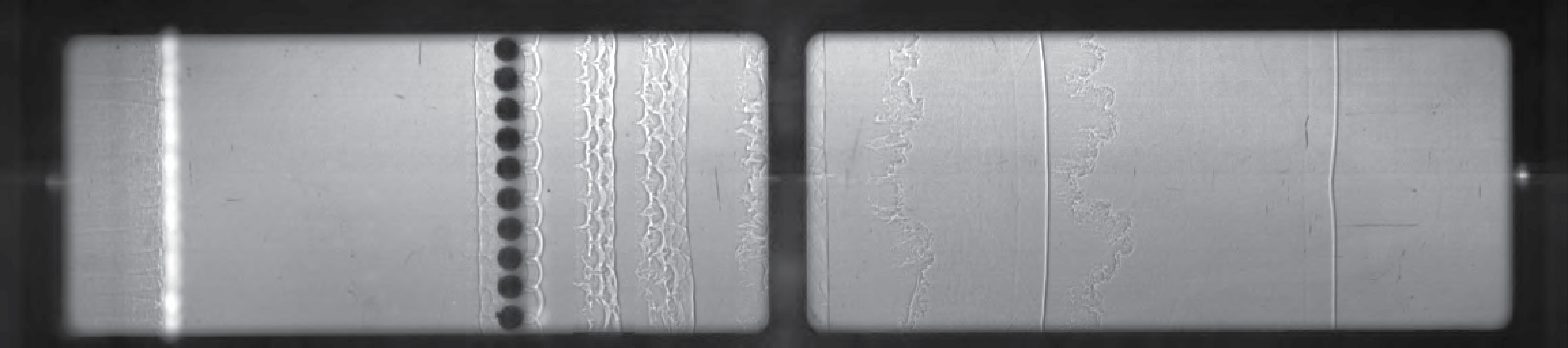}
\end{center}
\caption{Overlay of multiple shadowgraph video frames showing the evolution of the fast flame in a 2C$_2$H$_2$+5O$_2$+21Ar mixture at 9.6 kPa with a 75 \% blockage ratio; see video \ref{fig:acetylene75_9.6}.}
\label{fig:acetylene75_9.6}
\end{figure*}

For sufficiently low initial pressures, DDT was not observed, and the lead shock decayed substantially below the prediction made assuming a CJ deflagration.  Fig. \ref{fig:acetylene75_9.6} shows an example, which clearly shows how the localized ignition spots close to the row of cylinders are no longer replicated downstream.  The lead shock then separates by a large distance from the wrinkled flame.  In this case, the shock speed is well predicted by the model of Radulescu et al. \cite{Radulescuetal2015} assuming negligible energy release downstream of the row of cylinders, as shown in Fig. \ref{fig:speedsacetylene75}.  

\begin{figure*}
\begin{center}
\includegraphics[width=120mm]{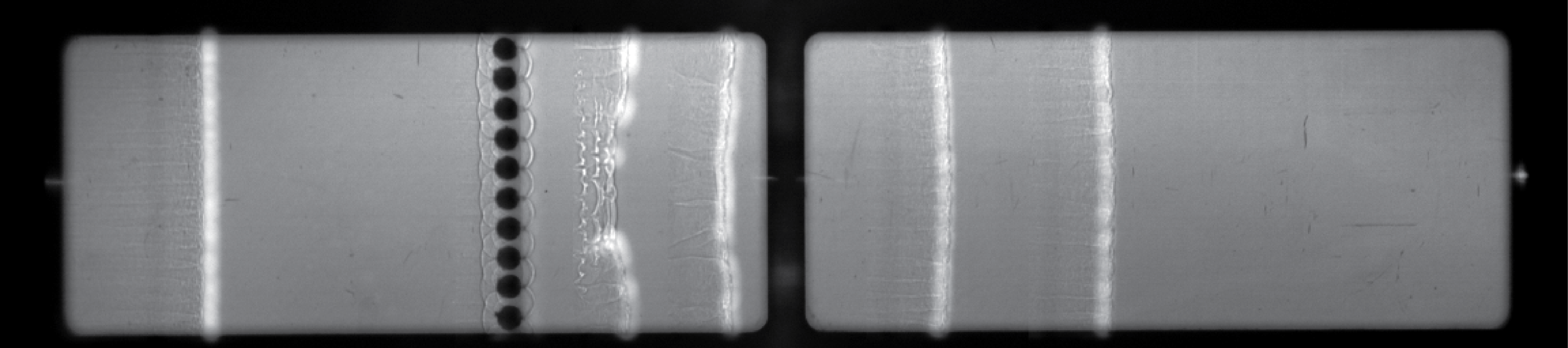}
\end{center}
\caption{Overlay of multiple shadowgraph video frames showing the evolution of the fast flame in a 2C$_2$H$_2$+5O$_2$+21Ar mixture at 11 kPa with a 75 \% blockage ratio; see video \ref{fig:acetylene75_11}.}
\label{fig:acetylene75_11}
\end{figure*}
Figures \ref{fig:acetylene75_11} and \ref{fig:acetylene75_9.6} show two examples of a successful DDT and decay to an inert shock, respectively (see also videos \ref{fig:acetylene75_11} and \ref{fig:acetylene75_9.6} provided as supplemental material).  For this mixture, it was found that the cylinders much more efficiently quenched the incident detonation, resulting in the DDT or failure occurring much closer to the cylinders.  Figure \ref{fig:acetylene75_9.6} shows the decoupled shock flame complex, with the lead shock rapidly recovering its planarity, followed by a conventional slow wrinkled flame.
\section{Discussion of main findings}
\addvspace{10pt}
The results obtained for the diluted acetylene mixtures, in the case of detonation failure of Fig. \ref{fig:acetylene75_9.6}, illustrate that the perturbations provided by the row of cylinders decay very fast, as first noted by Grondin and Lee \cite{Grondin&Lee2010} and computed by Radulescu et al. for the an inert interaction \cite{Radulescuetal2015}.  This observation substantiates the conclusion that the fast flames observed in all the other mixtures at comparable (physical) distances from the obstacles are supported by self-generated turbulence, as elegantly noted by Grondin and Lee \cite{Grondin&Lee2010}.   

For all the tests performed that resulted in DDT, the speed of the fast flame was found to be in excellent agreement with the model of Radulescu et al., which assumes a CJ deflagration.  It thus appears that a necessary condition for DDT is the establishment of a quasi-steady flame propagating at a sonic speed relative to the burned gases.  The DDT mechanism in all the experiments performed suggest that the flame-shock complex re-organizes into fewer stronger modes, as first observed by Radulescu et al. \cite{Radulescuetal2005}.   Possible mechanisms have been discussed by Radulescu and Maxwell \cite{Radulescu&Maxwell2011} and Maley et al.  \cite{Maleyetal2015}.  The structure and evolution of the fast flame appears identical to that of very unstable detonations, relying on turbulent mixing via hydrodynamic instabilities \citep{Radulescuetal2007}, powered by discete hotspots from triple point collisions.  Numerical reconstruction of the turbulent flow fields in such fast flames is currently underway in our laboratory, with preliminary results provided by Maxwell \cite{Maxwell2016}. 

Figure \ref{fig:Lddtcellsize} shows the variation of the DDT length for all the mixtures studied in the 75\% and 90\% blockage ratio configurations.  These are shown in terms of the $\chi$ parameter given by \eqref{eq:chi}, previously proposed as an indicator for the propensity of a reactive mixture to develop hotspots (see \cite{Radulescuetal2013} for review).  This parameter was calculated for each test mixture for conditions behind the lead shock supported by a CJ deflagration, as described in \cite{Radulescu2003}. Satisfactory correlations can be drawn for each set of data at the two blockage ratios, with typical correlations in the range of 7-20$\lambda$ for the 75 blockage ratio and 20-50$\lambda$ for the 90\% blockage ratio.  Furthermore, the data strongly support a decreasing relationship with increasing $\chi$ reflecting that a mixture with a stronger propensity for hotspots transits to a detonation faster.  Nevertheless, the significant differences between the two blockage ratio datasets indicate that no such correlation can be made universally. Slower DDT for higher blockage ratios is expected, as larger blockage ratios yield weaker transmitted shocks, which lead to longer chemical kinetic timescales. 
  
In order to reconcile the role of the characteristic chemical kinetic rates in the fast flames, we have normalized the DDT length by the characteristic distance $L_{1D}$, which is reflective of the induction delays behind the lead shock via \eqref{eq:L1D}.  The results are shown in Fig.\ \ref{fig:LddtL1D}.  The data fit approximately an inverse power law of the form
\begin{align}
L_{DDT}  \propto \frac{L_{1D}}{\chi} \propto c t_r \left( \frac{E_a}{RT} \right)^{-1}
\end{align}
The data reveal that a mixture with larger propensity for hotspot formation (larger $\chi$) is much more likely to undergo DDT under the same level of fluctuations.  Returning to the definition of $\chi$ given in \eqref{eq:chi}, with $L_{1D} \sim c t_i$, the inverse power law correlation suggests that the induction time drops out of the correlation for $L_{DDT}$. The sole chemical kinetic timescale of importance in the DDT process is the characteristic energy release time in a fluid particle.  This finding can be anticipated from Sharpe's treatment of DDT behind a lead shock in an individual hotspot \cite{Sharpe2002}, substantiated by Tang and Radulescu's results \cite{Tang&Radulescu2013}, which showed that the acceleration of an individual hotspot front is controlled by the inverse of this time scale, multiplied by the reduced activation energy.

\section*{Acknowledgements}

M.I.R acknowledges the financial support from both Shell and NSERC via a Collaborative Research and Development Grant, "Quantitative assessment and modeling of the propensity for fast flames and transition to detonation in methane, ethane, ethylene and propane" and the continuous inspiration from UgoBugo and BambaLeo.



\bibliographystyle{ieeetr}
\bibliography{references}







\end{document}